\documentclass[11pt] {article}

\usepackage{epsfig}
\catcode`@=11

\def\@citex[#1]#2{\if@filesw\immediate\write\@auxout{\string\citation{#2}}\fi
  \def\@citea{}\@cite{\@for\@citeb:=#2\do
    {\@citea\def\@citea{,\penalty\@m}\@ifundefined
      {b@\@citeb}{{\bf ?}\@warning
       {Citation `\@citeb' on page \thepage \space undefined}}%
\hbox{\csname b@\@citeb\endcsname}}}{#1}}

\def\citer{\@ifnextchar [{\@tempswatrue\@citexr}{\@tempswafalse\@citexr[]}}

\def\@citexr[#1]#2{\if@filesw\immediate\write\@auxout{\string\citation{#2}}\fi
  \def\@citea{}\@cite{\@for\@citeb:=#2\do
    {\@citea\def\@citea{--\penalty\@m}\@ifundefined
       {b@\@citeb}{{\bf ?}\@warning
       {Citation `\@citeb' on page \thepage \space undefined}}%
\hbox{\csname b@\@citeb\endcsname}}}{#1}}
\catcode`@=12
\title{  { \bf
Leading logarithmic QCD corrections  
to the $B_{s}\rightarrow\gamma\gamma$ decay rate including
long-distance effects through
$B_{s} \rightarrow \phi \gamma \rightarrow \gamma \gamma$ }}
\author{\vspace{1cm}\\
	{\bf G. Hiller}\thanks{E-mail address: ghiller@x4u2.desy.de} , \\
	 Deutsches Elektronen-Synchrotron DESY, Hamburg \\
	\vspace{5mm}\\
        {\bf E. O. Iltan}
        \thanks{E-mail address:
        eiltan@heraklit.physics.metu.edu.tr}
         \thanks{Supported by TUBITAK} \\
	Physics Department, Middle East Technical University \\
	Ankara, Turkey\\}
\date{}

\begin{document}
\setlength{\baselineskip}{24pt}

\maketitle
\begin{picture}(0,0)
       \put(325,340){DESY 97-076}
       \put(325,325){hep-ph/9704385}
       \put(325,310){April 1997}
\end{picture}
\vspace{-24pt}
\setlength{\baselineskip}{7mm}

\begin{abstract}
We present leading logarithmic QCD corrections to the decay 
$B_{s}\rightarrow\gamma\gamma$ in the Standard Model. 
Further, the form factor $F_{1}(0)$ of 
$B_{s}\rightarrow\phi\gamma$ is calculated
in the framework of QCD sum rules and found to be in agreement with the 
result existing in the literature. 
Using Vector Meson Dominance model, the amplitude for 
$B_s \rightarrow \phi \gamma \rightarrow\gamma\gamma$ is calculated
as an estimate of the $O_7$-type contributions to the long-distance effects 
in the $B_{s}\rightarrow\gamma\gamma$ decay.
The resulting branching ratio 
${\cal B}( B_s \rightarrow \gamma \gamma)_{SD+LD_{O_7}}$ 
is analysed in view of its strong dependence on the non-perturbative 
parameter $\bar{\Lambda}_s$, describing bound state effects, and the 
renormalization scale $\mu$. 
\end{abstract}

\thispagestyle{empty}
\newpage
\setcounter{page}{1}

\section{Introduction}
Rare B decays induced by flavor changing neutral currents (FCNC) are known to 
provide information about the Standard Model (SM) at quantum level and
quantitative information on the SM parameters, 
such as the Cabibbo-Kobayashi-Maskawa (CKM) matrix elements.
The CLEO observation \cite{cleo} of the radiative decay mode 
$B \rightarrow X_{s} \gamma $
has been analysed in the SM and the rate agrees with the SM-based
theoretical calculations \cite{nlo}.
Another example is $B_{s}\rightarrow \phi\gamma $, which is CKM allowed due 
to the dominant CKM matrix element dependence of the decay rate.
The calculational procedure of such decay rates is to use an effective 
Hamiltonian obtained by integrating out the top quark and the $W^{\pm}$
bosons \cite{effham} 
\begin{eqnarray}
{\cal{H}}_{eff}=-4 \frac{G_{F}}{\sqrt{2}} V_{tb} V^{*}_{ts} 
\sum_{i=1}^{8} C_{i}(\mu) O_{i}(\mu) \, \, .
\label{hamilton}
\end{eqnarray}
Here $O_{i}$ are suitable operators and $C_{i}$ are Wilson coefficients
renormalized at the scale $\mu$. The coefficients can be calculated 
perturbatively. Hadronic matrix elements $ <V|O_{i}|B>$ can be calculated
using some non-perturbative methods like 
QCD sum rules, which is one of the powerful methods to calculate matrix 
elements in a model independent way.

Among rare decays, $B_{s}\rightarrow \gamma \gamma$ is a potential 
candidate to test the SM and search for new physics. The final state contains 
CP-odd and CP-even states, 
allowing us to study CP violating effects.
Measurement of these odd and even states is a powerful test of the 
underlying theory, in particular SM. 
In the literature, $B_{s}\rightarrow \gamma \gamma$ decay has been 
investigated earlier in the lowest order 
\citer{yao,aliev}
and the branching ratio is found to 
be $4.5 \cdot 10^{-7}$ in the SM context for $m_s=0.5$ GeV and 
other parameters given in Table~\ref{parameters}, using the constituent quark 
model.

In the present work we give the leading logarithmic QCD-improved rates for
$B_s \rightarrow \gamma \gamma$. This can be achieved through a matching
of the full theory with the effective theory at a scale $\mu=m_W$,
using the effective Hamiltonian in eq.~(\ref{hamilton}),
and performing an evolution of the Wilson coefficients from 
$m_W$ down to $\mu \sim {\cal O}(m_b)$, thus 
resumming all large logarithms of the form 
$\alpha_s^n(m_b) log^m(\frac{m_b}{m_W})$, where $m \leq n$ 
($n=0,1,2, \ldots$). In the leading logarithmic approximation, which we use
here, $m=n$.
The effective Hamiltonian in eq.~(\ref{hamilton}) is identical  for 
$b \rightarrow s \gamma$ and for $b \rightarrow s \gamma \gamma$ to this order 
of $\frac{1}{m_W^2}$. Since there exists after applying the equations of 
motion no gauge-invariant FCNC-2-photon operator with field dimension $\leq 6$,
the set of operators (eq.~(\ref{o7})) is a basis for both 
decays \cite{grinstein}.
The bound state effects of the $B_s$ meson are modeled through an 
Heavy Quark Effective Theory (HQET) inspired approach following \cite{MW}.
We estimate further the additional contribution in the decay 
$B_{s}\rightarrow \gamma \gamma$ through $B_{s}\rightarrow \phi \gamma$
followed by $\phi\rightarrow \gamma$ using 
Vector Meson Dominance (VMD) \cite{terasaki}. 
In the language of the operator basis in eq.~(\ref{hamilton}), this 
contribution involves the operator $O_7$, (see eq.~(\ref{o7}) below).
The decay
$B_{s}\rightarrow \phi \gamma$ was studied in the literature in the framework 
of Light-cone QCD sum rules \cite{alibraunsimma}.
We have repeated the calculation using the ordinary QCD sum rules including
the contribution from the gluon condensate.
The CP-odd and CP-even amplitudes in $B_{s}\rightarrow \gamma \gamma$ are then 
estimated by considering the $\phi\rightarrow \gamma$ process using a 
$\phi$-photon conversion factor supplied by the VMD model.
In this part an extrapolation from $p'^2=m_{\phi}^2$
(needed for $B_s \rightarrow \phi \gamma$)  
to $p'^2=0$
(required for $B_s \rightarrow \gamma \gamma$) 
is necessary. 
We assume, that the form factor is dominated by a single pole,
which is a good approximation for light mesons. 
The decay rate for $B_{s}\rightarrow \gamma \gamma$ depends sensitively on the
model parameters $(m_b,\bar{\Lambda}_s)$ and $\mu$. For typical values
$(m_b,\bar{\Lambda}_s)=(5 \,\mbox{GeV},370 \, \mbox{MeV})$ and $\mu$=5 GeV,
we get (including long-distance effects through $O_7$) the branching ratio
${\cal B}(B_{s}\rightarrow \gamma \gamma))_{SD+LD_{O_7}}=1.18 \cdot 10^{-6}$,
which is a factor $1.9$ larger compared to the lowest order estimate for the 
same values of the parameters.
However, varying $(m_b,\bar{\Lambda}_s)$ and $\mu$ in the allowed range results
in significant variation on the branching ratio, yielding
$0.38 \cdot 10^{-6} \leq
{\cal B}(B_s \rightarrow \gamma \gamma)_{SD+LD_{O_7}} 
\leq 1.43 \cdot 10^{-6}$.

The paper is organized as follows:
In section 2 we display the amplitude for $B_{s}\rightarrow \gamma \gamma$ in 
a HQET inspired model and present the leading logarithmic QCD corrections. 
In section 3, we calculate the form factor $F_1$ in the decay
$B_{s}\rightarrow\phi\gamma$ 
using QCD sum rules and compare our result with the previous result
obtained in \cite{alibraunsimma}.
Section 4 is devoted to the estimate of the
$B_{s}\rightarrow \phi \gamma \rightarrow \gamma \gamma$ amplitude in the 
framework of VMD. We discuss the resulting branching ratio 
${\cal B}(B_{s}\rightarrow \gamma \gamma)_{SD+LD_{O_7}}$ and its
parametric dependence on the model parameters $(m_b,\bar{\Lambda}_s)$ 
and the scale $\mu$ in section 5.
\section{\bf Leading logarithmic improved short-distance contributions in 
$B_{s}\rightarrow \gamma \gamma$ decay}
\begin{figure}[htb]
\vskip -0.6truein
\centerline{\epsfysize=7in
{\epsffile{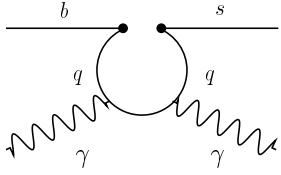}}}
\vskip -5.0truein
\caption[]{ The generic diagram contributing to $b \rightarrow s \gamma 
\gamma$ in the effective theory due to the (Fierz ordered) four-quark 
operators. 
The diagram with interchanged photons is not shown.}
\label{fig:o2}
\end{figure}

\begin{figure}[htb]
\vskip -0.6truein
\centerline{\epsfysize=7in
{\epsffile{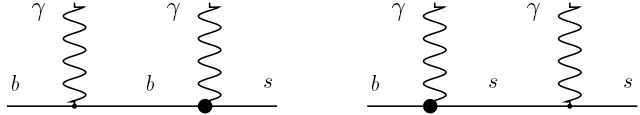}}}
\vskip -5.0truein 
\caption[]{ The reducible diagrams contributing to $b \rightarrow s \gamma 
\gamma$. The blob denotes the FCNC operator $O_7$. The diagrams with interchanged photons are not shown.}
\label{fig:o7}
\end{figure}

The amplitude for the decay $B_{s}\rightarrow \gamma \gamma$ can be 
decomposed as
\citer{yao,aliev}
\begin{equation}
{\cal A}(B_{s}\rightarrow \gamma \gamma)=
\epsilon_1^{\mu}(k_1) \epsilon_2^{\nu}(k_2)
(A^{+} g_{\mu \nu} +
i A^{-} \epsilon_{\mu \nu \alpha \beta} k_1^{\alpha}  k_2^{\beta}) \, \, ,
\label{am}
\end{equation}
where the $k_i$ and $\epsilon_i^{\nu}(k_i)$ denote the four-momenta and the 
polarization vectors of the outgoing photons, respectively 
\footnote{We adopt the convention 
$Tr(\gamma^{\mu} \gamma^{\nu} \gamma^{\alpha} \gamma^{\beta} \gamma_5)=
4i\epsilon^{\mu \nu \alpha \beta}$, with $\epsilon^{0 1 2 3}=+1$.}.
Using the effective Hamiltonian in eq.~(\ref{hamilton}),
the CP-even ($A^{+}$) and CP-odd ($A^{-}$) parts in the SM can be written 
as (for diagrams see fig.~\ref{fig:o2} and fig.~\ref{fig:o7})
in a HQET inspired approach
\footnote{In an earlier version of this paper the contributions of the
operators $O_{1,3 \dots 6}$ in the irreducible part in $A^{+}$ and $A^{-}$
were not completely taken into account. This is corrected here.
As a second improvement we give the amplitudes in a formalism inspired by HQET 
to estimate the uncertainties coming from the bound state.}
:
\begin{eqnarray}
A^{+}&=&-\frac{\alpha_{em} G_F}{\sqrt{2} \pi} f_{B_s} \lambda_t 
\left( \frac{1}{3}
\frac{m^4_{B_s} (m_b^{eff}-m_s^{eff})}{\bar{\Lambda}_s 
(m_{B_s}-\bar{\Lambda}_s) (m_b^{eff}+m_s^{eff})} 
C_7^{eff}(\mu)
\nonumber
\right.\\
&-&
\left.
\frac{4}{9} \frac{m_{B_{s}^2}}{m_b^{eff}+m_s^{eff}}
(-m_b J(m_b)+ m_s J(m_s) ) D(\mu) 
\right)  ,\nonumber\\
A^{-}&=&-\frac{\alpha_{em} G_F}{\sqrt{2} \pi} 2 f_{B_s} \lambda_t 
\left( \frac{1}{3}
\frac{1}{m_{B_s} \bar{\Lambda}_s (m_{B_s}-\bar{\Lambda}_s)} g_{-}
C_7^{eff}(\mu) \nonumber
\right.\\
&-&
\left.
\sum_q Q_q^2 I(m_q) C_q(\mu) +
\frac{1}{9 (m_b^{eff}+m_s^{eff})} 
(m_b \triangle(m_b)+m_s \triangle(m_s)) D(\mu)
\right) \, \, ,
\label{amplitudes}
\end{eqnarray}
where we have used the unitarity of the CKM-matrix 
$\sum_{i=u,c,t} V_{is}^{*} V_{ib}=0 $ 
and have neglected the contribution due to 
$V_{us}^{*} V_{ub} \ll V_{ts}^{*} V_{tb}\equiv \lambda_t$. 
In eq.~(\ref{amplitudes}) $N_{c}$ is the colour factor ($N_c=3$ for QCD)
and $Q_q=\frac{2}{3}$ for $q=u,c$ and $Q_q=-\frac{1}{3}$ for $q=d,s,b$.
The QCD-corrected Wilson coefficients 
in leading logarithmic approximation \cite{effham}, 
$C_{1 \dots 6}(\mu)$ and $C_7^{eff}(\mu)$,
enter the amplitudes in the combinations
\begin{eqnarray}
C_u(\mu)&=&C_d(\mu)=(C_3(\mu)-C_5(\mu)) N_c +C_4(\mu)-C_6(\mu) \, \, , 
\nonumber \\
C_c(\mu)&=&
(C_1(\mu)+C_3(\mu)-C_5(\mu)) N_c +C_2(\mu)+C_4(\mu)-C_6(\mu) \, \, ,
\nonumber \\
C_s(\mu)&=&C_b(\mu)=
(C_3(\mu)+C_4(\mu))(N_c+1)-N_c C_5(\mu)-C_6(\mu) \, \, , \nonumber \\
D(\mu)&=&C_5(\mu)+C_6(\mu) N_c \, \, .
\end{eqnarray} 
While $C_{1 \dots 6}(\mu)$ are the coefficients of the operators 
$O_{1 \dots 6}$, $C_7^{eff}(\mu)$ is the "effective" coefficient of $O_7$ and 
contains renormalization scheme dependent contributions from the 
four-quark operators
$O_{1\ldots 6}$ in ${\cal H}_{eff}$ to the effective vertex in
$b \rightarrow s \gamma$.
In the NDR scheme, which we use here, 
$C_7^{eff}(\mu)=C_7(\mu)-\frac{1}{3} C_5(\mu)-C_6(\mu)$, see \cite{effham} 
for details.
The initial values of $C_{1 \dots 6}(\mu)$ and $C_7^{eff}(\mu)$ in the SM are
\begin{eqnarray}
C_{1,3 \dots 6}(m_W)&=&0 \nonumber \, \, , \\
C_2(m_W)&=&1 \nonumber \, \, , \\
C_7^{eff}(m_W)&=&\frac{3 x^3-2 x^2}{4(x-1)^4} \ln x+
\frac{-8x^3-5 x^2+7 x}{24 (x-1)^3} \, \, ,
\end{eqnarray}
and $x=m_t^2/m_W^2$.
For comparison, $C_1(m_b)=-0.246$, $C_2(m_b)=1.106$, 
$C_3(m_b)=0.011$, $C_4(m_b)=-0.025$, $C_5(m_b)=0.007$, $C_6(m_b)=-0.031$ and 
$C_7^{eff}(m_b)=-0.313$ for 
the input values given in Table~\ref{parameters}.
The operator basis of ${\cal{H}}_{eff}$ is given as
\begin{eqnarray}
 O_1 &=& (\bar{s}_{L \alpha} \gamma_\mu b_{L \alpha})
               (\bar{c}_{L \beta} \gamma^\mu c_{L \beta}), \nonumber   \\
 O_2 &=& (\bar{s}_{L \alpha} \gamma_\mu b_{L \beta})
               (\bar{c}_{L \beta} \gamma^\mu c_{L \alpha}),  \nonumber   \\
 O_3 &=& (\bar{s}_{L \alpha} \gamma_\mu b_{L \alpha})
               \sum_{q=u,d,s,c,b}
               (\bar{q}_{L \beta} \gamma^\mu q_{L \beta}),  \nonumber   \\
 O_4 &=& (\bar{s}_{L \alpha} \gamma_\mu b_{L \beta})
                \sum_{q=u,d,s,c,b}
               (\bar{q}_{L \beta} \gamma^\mu q_{L \alpha}),   \nonumber  \\
 O_5 &=& (\bar{s}_{L \alpha} \gamma_\mu b_{L \alpha})
               \sum_{q=u,d,s,c,b}
               (\bar{q}_{R \beta} \gamma^\mu q_{R \beta}),   \nonumber  \\
 O_6 &=& (\bar{s}_{L \alpha} \gamma_\mu b_{L \beta})
                \sum_{q=u,d,s,c,b}
               (\bar{q}_{R \beta} \gamma^\mu q_{R \alpha}),  \nonumber   \\  
 O_7 &=& \frac{e}{16 \pi^2}
          \bar{s}_{\alpha} \sigma_{\mu \nu} (m_b R + m_s L) b_{\alpha}
                F^{\mu \nu},                             \nonumber       \\
 O_8 &=& \frac{g}{16 \pi^2}
    \bar{s}_{\alpha} T_{\alpha \beta}^a \sigma_{\mu \nu} (m_b R + m_s L)  
          b_{\beta} G^{a \mu \nu},  
\label{o7}
\end{eqnarray}
where $L$ and $R$ denote chiral projections, $L(R)=1/2(1\mp \gamma_5)$ and
$\alpha$ and $\beta$ are $SU(3)$ colour indices.
Note that $O_8$ does not contribute here in this order of $\alpha_s$.
The functions $I(m_q), \, J(m_q)$ and $\triangle(m_q)$ come from the 
irreducible diagrams with an internal 
$q$ type quark propagating, see fig.~\ref{fig:o2}, and are defined as
\begin{eqnarray}
I(m_q)&=&1+\frac{m_q^2}{m_{B_s}^2} \triangle (m_q) \, \, , \nonumber \\
J(m_q)&=&1-\frac{m_{B_s}^2-4 m_q^2}{4 m_{B_s}^2} \triangle(m_q)  \, \, ,
\nonumber \\
\triangle(m_q)&=&\left(
\ln(\frac{m_{B_s}+\sqrt{m_{B_s}^2-4 m_q^2}}
         {m_{B_s}-\sqrt{m_{B_s}^2-4 m_q^2}})-i \pi \right)^2 
\, \,{\mbox{for}}\, \, \frac{m^2_{B_s}}{4 m_q^2} \geq 1 , \nonumber \\
\triangle(m_q)&=&-\left(
2 \arctan(\frac{\sqrt{4 m_q^2-m_{B_s}^2}}
         {m_{B_s}})- \pi \right)^2 
\, \,{\mbox{for}}\, \, \frac{m^2_{B_s}}{4 m_q^2} <\ 1 .
\end{eqnarray}
The parameter $\bar{\Lambda}_s$ enters eq.~(\ref{amplitudes}) through the 
bound state kinematics. 
For definiteness, we consider the decay 
$B_s\equiv (\bar{b} s) \rightarrow \gamma \gamma$.
We write the momentum of the $\bar{b}$-quark inside the meson as
$p=m_b v+k$, where $k$ is a small residual momentum, $v$ is the 4-velocity,
which connects the quark with the meson kinematics through $P=m_{B_s} v$
and $P$ is the momentum of the meson. In the $B_s$ rest frame, $v=(1,0,0,0)$.
For the reducible diagrams, see fig.~\ref{fig:o7}, we need to evaluate
$p.k_i$ and $p'.k_i$, $i=1,2$, where $k_i,p'$ are the momenta of the outgoing
photon and $s$-quark, respectively.
Now following \cite{MW}, we average the residual momentum of the 
$\bar{b}$-quark through 
\begin{eqnarray}
<k_{\alpha}>&=&
-\frac{1}{2 m_b} (\lambda_1+3 \lambda_2) v_{\alpha}\, \, ,\nonumber\\
<k_{\alpha} k_{\beta}>&=&\frac{\lambda_1}{3} 
(g_{\alpha \beta } -v_{\alpha} v_{\beta}) \, \, ,
\end{eqnarray}
where $\lambda_1, \lambda_2$ are matrix elements from the heavy quark 
expansion. Using $P=p-p'$, $P.k_i=\frac{m^2_{B_s}}{2}$,  
$v.k_i=\frac{m_{B_s}}{2}$ and the HQET relation \cite{MW}
\begin{equation}
m_{B_s}=m_b +\bar{\Lambda}_s-\frac{1}{2 m_b} (\lambda_1+3 \lambda_2)
\label{hqe}
\end{equation}
one gets:
\begin{eqnarray}
p.k_i &=&\frac{m_{B_s}}{2} (m_{B_s}-\bar{\Lambda}_s) \, \, ,\nonumber \\
p'.k_i&=&-\frac{m_{B_s}}{2} \bar{\Lambda}_s \, \, ,\nonumber \\
(m_b^{eff})^2\equiv p^2&=&m_b^2-3 \lambda_2\, \, ,\nonumber \\
(m_s^{eff})^2\equiv p'^2&=&(m_b^{eff})^2
-m^2_{B_s}+2 m_{B_s} \bar{\Lambda}_s  \, \, .
\label{kin}
\end{eqnarray}
The non-perturbative parameter $\bar{\Lambda}_s$ can be related to 
$\bar{\Lambda}$, which 
has been extracted (together with $\lambda_1$) from data on semileptonic 
$B^{\pm},B^0$ decays 
by \cite{gremm}, and the measured mass
difference $\triangle m=m_{B_s}-m_B=90$ MeV \cite{PDG}, defining
$\bar{\Lambda}_s=\bar{\Lambda}+\triangle m$.
The matrix element $\lambda_2$ is well determined from the 
$B^{\ast}_{(s)}-B_{(s)}$
mass splitting, $\lambda_2=0.12 \, {\mbox{GeV}}^2$.
With the help of eq.~(\ref{hqe}), the correlated values of $\bar{\Lambda}$  
and $\lambda_1$ can be transcribed into a 
correlation between $\bar{\Lambda}_{(s)}$ and $m_b$.
We select 3 representative values
\footnote{We choose 
$(\lambda_1, \bar{\Lambda})=(-0.09,280),(-0.19,390),(-0.29,500)$ in 
$( {\mbox{GeV}}^2, {\mbox{MeV}} )$ from fig.~1 in \cite{gremm}. }
$(m_b,\bar{\Lambda}_s)=(5.03,370),(4.91,480),(4.79,590)$ in 
$( {\mbox{GeV}}, {\mbox{MeV}} )$
to study the hadronic
uncertainties of our approach. 
Furthermore, we have used the definition 
\begin{eqnarray}
<0|\bar{s} \gamma_{\mu} \gamma_5 b|B_{s}(P)>&=&i f_{B_s} P_{\mu} \, \, ,
\end{eqnarray}
which leads together with
the off-shellness of the quarks inside the meson to the
matrix element of the pseudoscalar current 
\begin{eqnarray}
<0|\bar{s} \gamma_5 b|B_{s}(P)>&=&
-i f_{B_s} \frac{m^2_{B_s}}{m_b^{eff}+m_s^{eff}} \, \, .
\end{eqnarray}
The auxiliary function
$g_{-}=g_{-}(m_b^{eff},\bar{\Lambda}_s)$ is defined as
\begin{eqnarray}
g_{-}=m_{B_s}(m_b^{eff}+m_s^{eff})^2+
\bar{\Lambda}_s (m^2_{B_s}-(m_b^{eff}+m_s^{eff})^2) 
\,\, .
\end{eqnarray}
Note that in the limit $\bar{\Lambda}_s \to m_s$, 
$m_{b,s}^{eff} \to m_{b,s}$ and using $m_{B_s}=m_b+m_s$ we recover the 
result obtained by the constituent quark model \citer{yao,aliev},
ignoring QCD corrections.
Using the above expressions, the partial decay width is then given by :
\begin{equation}
\Gamma(B_{s}\rightarrow \gamma \gamma)=\frac{1}{32 \pi m_{B_s}} 
(4 |A^{+}|^2+\frac{1}{2} m_{B_s}^4|A^{-}|^2) \, \, .
\label{br}
\end{equation}

Now, there are 2 new observations to be made:\\
First, the Wilson coefficients in eq.~(\ref{amplitudes}) depend on the scale 
$\mu$. Therefore, since the behaviour of these short-distance (SD) 
coefficients under renormalization is known from the studies of
$B \rightarrow X_s \gamma $ \cite{nlo,effham}, one can give an improved width 
for $B_s \rightarrow \gamma \gamma$
by including the leading logarithmic QCD corrections, by 
renormalizing the coefficients $C_{1\dots 6}$ and $C_7^{eff}$ from 
$\mu=m_W$ down to the relevant scale $\mu \approx {\cal O}(m_b)$.
The explicit ${\cal O}(\alpha_s)$ improvement in the decay width
$\Gamma(B_s \rightarrow \gamma \gamma)$ requires the calculation of a 
large number of virtual corrections, which we have not taken into account.
Varying the scale $\mu$ in the range
$\frac{m_b}{2} \leq \mu \leq 2 m_b$,
one introduces an uncertainty, which can be reduced only when the complete 
next-to-leading order (NLO)-analysis is available, similar to the recently 
completed calculation for 
the $B \rightarrow X_s \gamma$ decay \cite{nlo}. 
%
%

The second point concerns the strong dependence of the decay width
$\Gamma(B_s \rightarrow \gamma \gamma)$
on $\bar{\Lambda}_s$,
$\Gamma \sim {\cal{O}}(\frac{1}{\bar{\Lambda}_s^2})$ in eq.~(\ref{br}). 
It originates in the $s$-quark propagator in the diagram with an intermediate 
$s$-quark in fig.~\ref{fig:o7}.
In the earlier 
work the authors of e.g. \cite{yao} evaluated the decay width with 
$m_s \approx m_K$, assuming that the constituent quarks are to be treated as 
static quarks in the meson.
This is a questionable assumption. 
In the HQET inspired approach, this gets replaced by $\bar{\Lambda}_s$, which 
is well-defined experimentally. This formalism implies, that the decay width
$\Gamma(B_d \rightarrow \gamma \gamma)$ will involve the parameter
$\bar{\Lambda}$, which avoids the unwanted uncertainty on $m_d$.

Lowering the scale $\mu$ from $\mu=m_W$ to $\mu \simeq {\cal O}(m_b)$ and 
$\bar{\Lambda}_s$
enhances the branching ratio ${\cal B}(B_{s}\rightarrow \gamma \gamma)$. 
The dependence of the branching ratio as a function of the scale $\mu$ for 
different values of $(m_b,\bar{\Lambda}_s)$ is discussed in the last 
section including the
$O_7$-type long-distance (LD) estimate. 
\begin{table}[h]
	\begin{center}
	\begin{tabular}{|l|l|}
	\hline
	\multicolumn{1}{|c|}{Parameter}	& 
		\multicolumn{1}{|c|}{Value}	\\
	\hline \hline
	$m_c$			& $1.4$ (GeV) \\
	$m_b$			& $4.8$ (GeV) \\
	$\alpha_{em}^{-1}$	& 129		\\
        $\lambda_t$            & 0.04 \\
	$\Gamma_{tot}(B_s)$		& $4.09 \cdot 10^{-13}$ (GeV) 	\\
        $f_{B_s}$             & $0.2$ (GeV)  \\   
        $m_{B_s}$             & $5.369$ (GeV) \\
        $m_{t}$             & $175$ (GeV) \\
        $m_{W}$             & $80.26$ (GeV) \\
        $m_{Z}$             & $91.19$ (GeV) \\
        $\Lambda^{(5)}_{QCD}$             & $0.214$ (GeV) \\
        $\alpha_{s}(m_Z)$             & $0.117$  \\
        $\lambda_2$             & $0.12$ $(\mbox{GeV}^2)$ \\
	\hline
	\end{tabular}
	\end{center}
\caption{Values of the input parameters used in the numerical
          calculations unless otherwise specified.}
\label{parameters}
\end{table}
\section{QCD sum rule for 
          the $B_{s}\rightarrow\phi\gamma$ form factor}
\subsection{Calculation of the sum rule}
The amplitude for the  $B_{s}\rightarrow\phi\gamma$  transition 
${\cal{A}}(B_{s}\rightarrow \phi\gamma)= <\phi\gamma|{\cal{H}}_{eff}|B_{s}>$
reduces to
\begin{eqnarray}
{\cal{A}}(B_{s}\rightarrow \phi\gamma)
=\epsilon^{\mu} C m_b <\phi(p')|\bar{s}\sigma_{\mu\nu} R q^{\nu}b|B_{s}(p)>
\label{amphigam}
\end{eqnarray}
with the constant C 
\begin{eqnarray}
C=\frac{G_{F}}{\sqrt{2}}\frac{e}{2 \pi^2} V_{ts}^{*} V_{tb} C^{eff}_{7}(\mu)\, \, ,
\label{const}
\end{eqnarray}
where
we just take the contribution due to the electromagnetic penguin 
operator $O_7$ into account and put $m_s=0$, justified by
 $m_s \ll m_b$.
Here $\epsilon$ and q are the photon polarization and the (outgoing) photon 
momentum, respectively.
Lorentz decomposition gives further:
\begin{eqnarray}
<\phi(p')|\bar{s}\sigma_{\mu\nu} R q^{\nu}b|B_{s}(p)>&=&i \epsilon_{\mu\nu\rho
\sigma} \epsilon^{\phi \nu} p^{\rho} p'^{\sigma} F_{1}(q^{2})\nonumber\\
&+& (\epsilon^{\phi}_{\mu} p.q-p_{\mu} q.\epsilon^{\phi})G(q^{2}) \, \, ,
\label{f1g}
\end{eqnarray}
where $p, \, p'$ denote the four-momenta of the initial $B_s$-meson and the 
outgoing $\phi$, respectively and $\epsilon_{\mu}^{\phi}$ is the polarization 
vector of the $\phi$-meson.
At the point $q^{2}=0$, it is enough to calculate $F_{1}(0)$,
since both form factors coincide \cite{alibraun}.
Note, that the form factors introduced above are in general functions of two 
variables $q^2$ and $p'^2$. Since $\phi$ is on-shell, we abbreviate here 
and in the following unless otherwise stated
$F_1(q^2) \equiv F_1(q^2, p'^2=m_{\phi}^2)$.

\begin{figure}[htb]
\vskip -0.5truein
\centerline{\epsfysize=12cm
{\epsffile{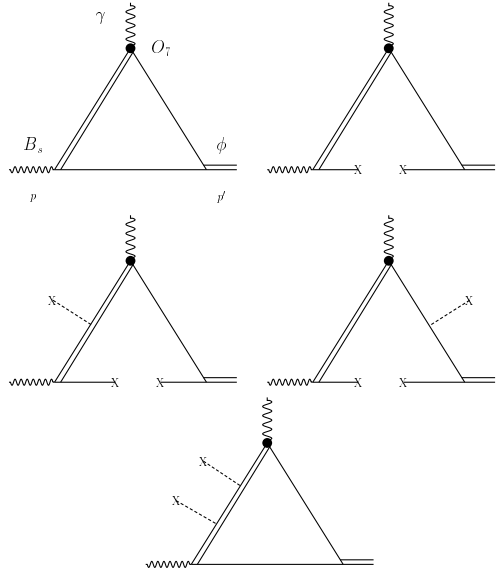}}}
\vskip -0.9truein
\caption[]{
Contributions of perturbation theory and of vacuum condensates to the 
$B_{s}\rightarrow\phi\gamma$ decay. The dashed lines denote soft gluons.}
\label{fig:diagrams}
\end{figure}

The starting point for the sum rule is the three-point 
function \cite{colangelo}
\begin{eqnarray}
T_{\alpha \mu}&=& -\int d^{4}x e^{ipx-ip'y}
<0|T[J_{\alpha} (x)T_{\mu}(0)J_{5}(y)]|0> \, \, ,
\label{talfabeta}
\end{eqnarray}
where $J_{\alpha}=\bar{s}\gamma_{\alpha}s$, $J_{5}=\bar{s}i\gamma_{5}b$ and
$T_{\mu}=\bar{s} \frac{1}{2} \sigma_{\mu\nu} q^{\nu} b$ correspond to the electromagnetic,
pseudoscalar currents and the penguin operator, respectively. 
Performing now an operator product expansion (OPE) of $T_{\alpha \mu}$, we 
obtain a perturbative term, the so-called bare loop, and non-perturbative power corrections, diagrammatically shown in fig.~\ref{fig:diagrams}.
The bare loop diagram can be obtained using a double dispersion relation in
$p^{2}$ and $p'^2$,
\begin{eqnarray}
T_{bare}=\frac{1}{\pi^2} \int_{m_b^2}^{\infty} ds \int_0^{\infty} ds' 
\frac{\rho(s,s')}{(s-p^2)(s'-p'^2)}
+ {\mbox{subtractions}} \, \, .
\label{bare}
\end{eqnarray}
Technically, the spectral density $\rho(s,s')$ can be calculated by
using the Cutkowsky rule, namely, by replacing the usual propagator 
denominator by a delta function: \\
$\frac{1}{k^{2}-m^{2}} \rightarrow -2\pi i \delta(k^{2}-m^2) \theta(k_0)$. 
As a result we get 
\begin{eqnarray}
\rho(s,s')=\frac{N_{c}}{8} m_{b}^{4}\frac{s'}{(s-s')^3} \, \, .
\label{rho}
\end{eqnarray}

OPE enables us further to parametrize the non-perturbative effects in 
terms of vacuum expectation values of gauge-invariant operators up to a 
certain dimension,  the so-called condensates.
We consider up to dimension-5 operators; i.e. the quark 
condensate, gluon condensate and the quark-gluon (mixed) condensate 
contributions 
(fig.~\ref{fig:diagrams}). 
This calculation
is carried out in the fixed point gauge, i.e. $A_{\mu}.x_{\mu}=0$. 
We get
\begin{eqnarray}
T_{dim-3}&=& \frac{-m_{b}}{2}
<\bar{s}s>\frac{1}{(p^{2}-m_{b}^{2}) p'^{2}}  \, \, ,\nonumber\\ 
T_{dim-4}&=&\frac{\alpha_s}{144 \pi} <G^2> \int_{0}^{1} dx \int_{0}^{1-x} dy
\int_{0}^{\infty} d \alpha \alpha^3 \nonumber \\ 
&\cdot& ( c_1 + c_2 P^2 + c_3 P'^{2}) e^{-\alpha (d_1 + d_2 P^2 +d_3 P'^2)}
\, \, , \nonumber \\
T_{dim-5}&=&\frac{m_{b}}{2}g<\bar{s}\sigma G s> [ \frac{m_{b}^{2}}
{2 (p^{2}-m_{b}^{2})^{3} p'^{2}}+\frac{m_{b}^{2}}{3
(p^{2}-m_{b}^{2})^{2} p'^{4}}
\nonumber \\ &+&\frac{1}{2 (p^{2}-m_{b}^2)^{2} p'^{2}} ] \, \, ,
\label{Ti}
\end{eqnarray}
where
%
%
\begin{eqnarray}
c_1&=&m_b^4 x^4  \, \, , \nonumber \\
c_2&=&m_b^2 x^4 (1 - x -y )  \, \, ,  \nonumber \\
c_3&=&m_b^2 x^3 (3+y) (1 - x - y)  \, \, ,  \nonumber \\
d_1&=&m_b^{2} x  \, \, ,  \nonumber \\
d_2&=&x (1 - x - y)  \, \, ,  \nonumber \\
d_3&=&y (1 - x - y ) \, \, .
\label{cidi}
\end{eqnarray}
Here we used the exponential representation for the gluon condensate
contribution:
\begin{equation}
\frac{1}{D^n}=\frac{1}{(n-1)!} \int_{0}^{\infty} d\alpha \, \, \alpha^{n-1}
e^{-\alpha D} \, \, .
\end{equation}
The momenta $P, \, P' $ in eq.~(\ref{Ti}) are euclidean.

For the calculation of the physical part of the sum rules we insert a complete set of on-shell states with the same quantum numbers as  $B_{s}$ and $\phi$ in
eq.~(\ref{talfabeta}) and get a double dispersion relation
\begin{eqnarray}
T_{phys}=\frac{m_{B_s}^2 f_{B_s}} {m_{b}} f_{\phi} m_{\phi}
\frac{1}{(p^{2}-m_{B_s}^2)(p'^{2}-m_{\phi}^2)} 
F_{1}(0)+ {\mbox{continuum}} \, \, ,
\end{eqnarray}
where $f_{\phi}$ and $f_{B_s}$ are the leptonic decay constants of the $\phi$ 
and $B_{s}$ mesons respectively, defined as usual by
\begin{eqnarray}
<0|J_{\alpha}|\phi>&=&m_{\phi} f_{\phi}
\epsilon_{\alpha}^{\phi}  \, \, , \nonumber \\ 
<0|J_{5}|B_{s}(p)>&=&f_{B_s} m_{B_s}^{2}/m_{b} \, \, .
\end{eqnarray}
We have absorbed all higher order states and resonances in the continuum.

Now, we equate the hadron-world with 
the quark-world by $T_{phys}=T_{bare}+T_3+T_4+T_5$. 
Using quark-hadron duality, we model the continuum contribution by purely 
perturbative QCD. To be definite, it is the part in eq.~(\ref{bare})
above the so-called continuum thresholds $s_0$ and $s'_{0}$.
To get rid of subtractions and to 
suppress the contribution of higher order states, we apply a Double Borel 
transformation $\hat{B}$ \cite{shifman} with respect to $p^2$ and $p'^{2}$. 
We make use of the following properties of the Borel transform:
\begin{eqnarray}
\hat{B}(\frac{1}{(p^2-m^2)^n})&=&\frac{(-1)^n}{(n-1)!} 
\frac{e^{-m^2/M^2}}{(M^2)^n}  \, \, , \\
\hat{B}(e^{-\alpha p^2})&=& \delta(1-\alpha M^2) \, \, .
\end{eqnarray}
Finally, this yields the sum rule:
\begin{eqnarray}
F_{1}(0)&=&
\exp(\frac{m_{B_s}^{2}}{M^{2}}+\frac{m_{\phi}^{2}}{M'^{2}}) \frac{m_{b}}
{f_{B_s} f_{\phi} m_{\phi} m_{B_s}^{2}} \{ \frac{1}{ \pi^{2}}
\int_{m_{b}^{2}}^{s_{0}} ds \int_{0}^{\bar{s}} ds' \rho (s,s')
e^{-s/M^{2}-s'/M'^{2}} \nonumber\\
&-&  \frac{m_{b}}{2} <\bar{s} s> e^{(-m_{b}^{2}/M^{2})} [1-
m_{0}^2(\frac{m_{b}^{2}}{4 M^{4}}+\frac{m_{b}^{2}}{3 M^{2} M'^{2}}-
\frac{1}{2 M^{2}} ) ] 
\nonumber \\
&+& \frac{\alpha_s}{\pi} <G^2> \int_{0}^{x_{max}} N(x) dx 
\} \, \, ,
\label{sumrule}
\end{eqnarray}
where $\bar{s}=min(s-m_{b}^{2},s'_{0})$
and $x_{max}=\frac{M'^{2}}{M^2+M'^{2}}$. 
Here we used the parametrization
\begin{eqnarray}
g<\bar{s} \sigma G s>&=m_{0}^{2} <\bar{s} s> \, \, .
\end{eqnarray}
\begin{figure}[htb]
\vskip -0.6truein
\centerline{\epsfysize=12.0cm
{\epsffile{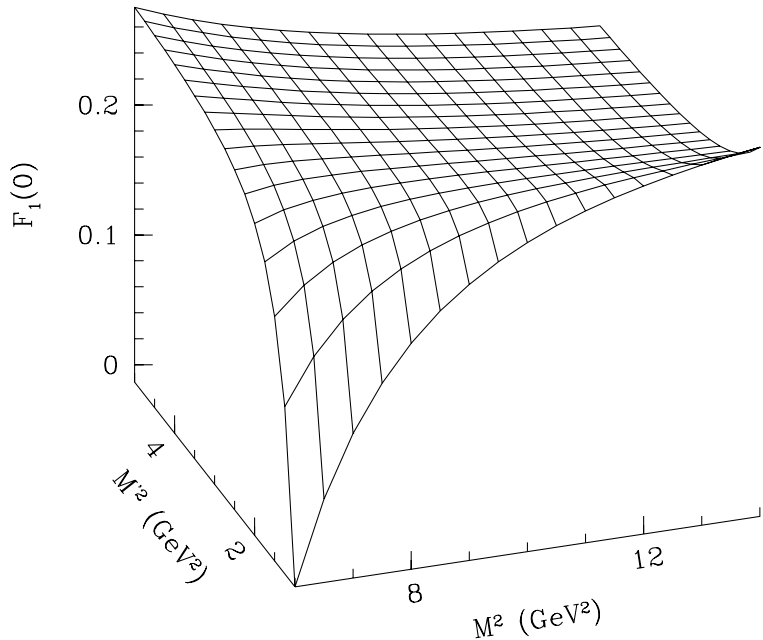}}}
\vskip -0.6truein
\caption[]{ The dependence of the decay constant $F_{1}(0)$
on the Borel parameters $M^{2}$ and $M'^{2}$ for 
$s_{0}=33\,\, {\mbox{GeV}}^{2}$.}
\label{fig:3d}
\end{figure}
\begin{figure}[htb]
\vskip -2.0truein
\centerline{\epsfysize=12.0cm
{\epsffile{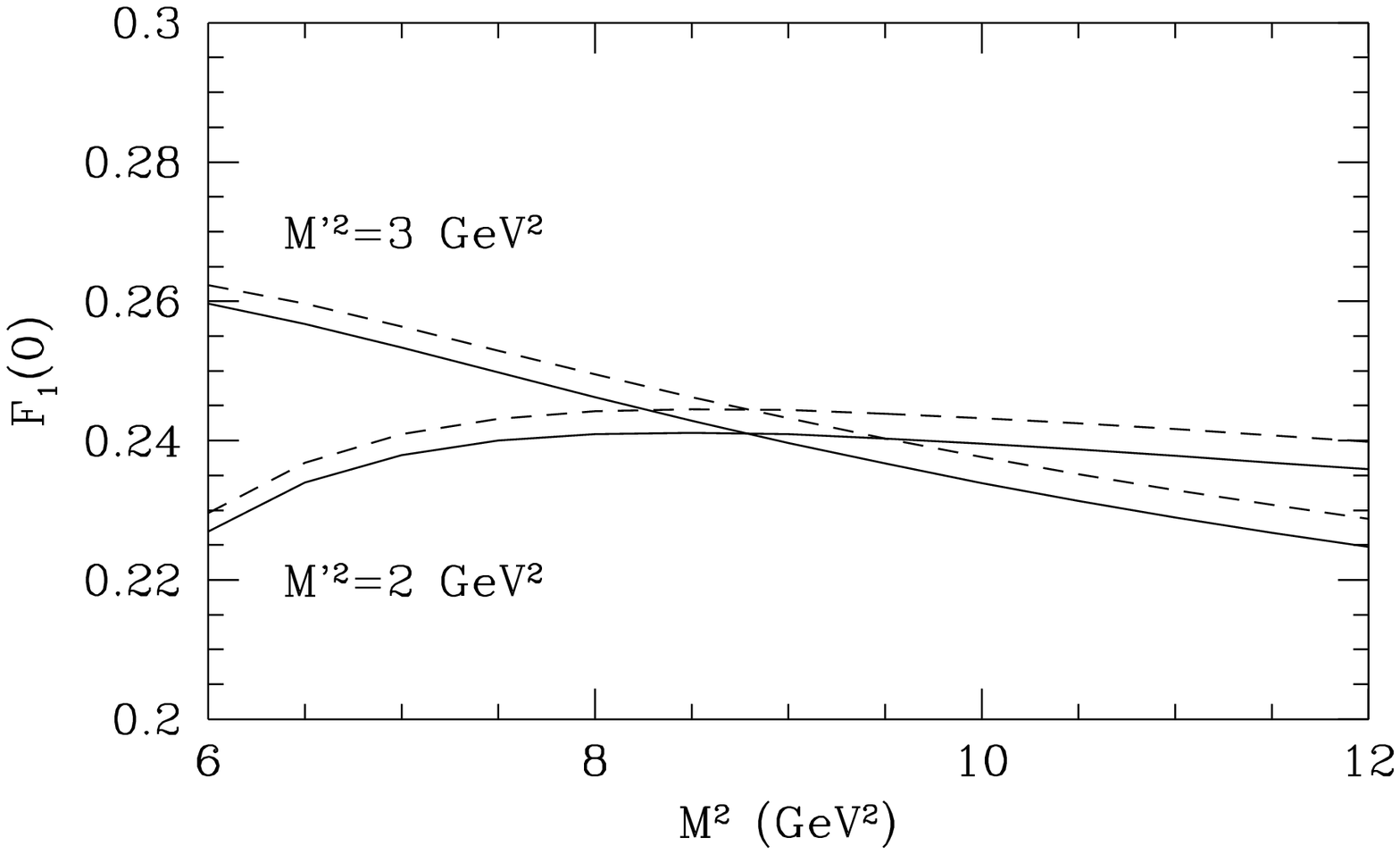}}}
\vskip -0.3truein
\caption[]{ The dependence of the decay constant $F_{1}(0)$
on the Borel parameter $M^{2}$ for fixed $M'^{2}$ at 
$s_{0}=33\,\,{\mbox{GeV}}^{2}$ (solid) and 
$s_{0}=35\,\,{\mbox{GeV}}^{2}$ (dashed). }
\label{fig:2d}
\end{figure}
The last term in eq.~(\ref{sumrule}) is
due to the gluon condensate contribution and the function $N(x)$ is defined
by:
\begin{eqnarray}
N(x)&=&\frac{1}{48} \exp(-\frac{m_b^2}
{M^2 (1 - x- x M^2/M'^{2})}) m^2_{b} M'^{6} x
(m^2_b M'^{4} - 4 M^2 M'^{4}+5 M^2 M'^{4} x \nonumber \\
&+&5 M^4 M'^{2} x- M^2 M'^{4} x^2 -2 M^4 M'^{2} x^2 - M^6 x^2)
/ (M^{4}(-M'^{2} + M'^{2} x + M^{2} x)^5) \, .
\end{eqnarray}

\subsection{Analysis of the sum rule}
First we list the values of the input parameters entering the sum rules
(eq.~(\ref{sumrule})), which are not included in Table \ref{parameters}: 
$m_{0}^{2}=0.8\,\,{\mbox{GeV}}^{2} $ \cite{mzero},
$<\bar{s}s>=-0.011 \, \, {\mbox{GeV}}^{3}$ \cite{sbars}, 
$\frac{\alpha_{s}}{\pi}<G^{2}>=0.03 \,\, {\mbox{GeV}}^{4}$ \cite{shifman}, 
$m_{\phi}=1.019 \,\,{\mbox{GeV}}$ and 
$f_{\phi}=0.23 \, \, {\mbox{GeV}}$ \cite{chernyak}.

We do the calculations for two different continuum threshold values
$s_{0}=33 \,\,{\mbox{GeV}}^2$ and $s_{0}=35\,\,{\mbox{GeV}}^2$ and take 
$s'_{0}=1.8 \,\, {\mbox{GeV}}^{2}$.
In fig.~\ref{fig:3d} we present the dependence of $F_{1}(0)$
on $M^{2}$ and $M'^{2}$ for $s_{0}=33 \,\, {\mbox{GeV}}^{2}$.
According to the QCD sum rules method,
it is necessary to find a range of $M^{2}$ and $M'^{2}$, where the
dependence of $F_{1}(0)$ on these parameters is very weak and, at the
same time, the power corrections and the continuum contribution remain
under control. 
{}From fig.~\ref{fig:3d} and fig.~\ref{fig:2d} follows that the best
stability region for $F_{1}(0)$ is 
$7\,\,{\mbox{GeV}}^{2}\leq M^{2}\leq 9\,\,{\mbox{GeV}}^{2}$,
$2\,\,{\mbox{GeV}}^{2}\leq M'^{2}\leq 3\,\,{\mbox{GeV}}^{2}$ for 
$s_{0}=33,35\,\,{\mbox{GeV}}^{2}$. We get:
\begin{eqnarray}
F_{1}(0)= 0.24 \pm 0.02 \, \, .
\end{eqnarray}
This agrees for our value of $m_b$ within errors with the result
given in the literature, 
based on Light-cone QCD sum rule calculations \cite{alibraunsimma}.

Numerical analysis shows, as also mentioned in \cite{colangelo}, that the 
natural hierarchy of the bare loop, the power corrections and 
continuum contributions does not hold due to the smallness of the integration 
region, and the power corrections exceed the bare loop contribution.
The gluon condensate contribution is $\leq 1 \% $ of
the dim-3 $+$ dim-5 condensate contributions 
and can therefore be safely neglected in numerical calculations.

\section{The $B_{s} \rightarrow \phi\gamma\rightarrow \gamma \gamma$ amplitude
using VMD model}
We consider the construction of a VMD amplitude using 
the amplitude for the decay
$B_{s}\rightarrow\phi\gamma$ as an input. Our aim is to continue the 
$B_{s}\rightarrow\phi\gamma$ decay amplitude from $p'^{2}=m_{\phi}^{2}$ 
to $p'^{2}=0$, such that the $\phi$ meson propagates as a massless virtual
particle before converting into a photon.
Note that we suppressed in our notation the dependence of the form factor
$F_1(q^2)=F_1(q^2,p'^2= m_{\phi}^2)$ on the second argument $p'^2$. 
We define here
$\bar{F_1}(Q^2) \equiv F_1(q^2=0,Q^2)$ for virtual momenta $Q^{2}=-p'^{2}$.
Assuming pole-type behaviour of the form factor $\bar{F_1}(Q^2)$ 
we extrapolate using the single-pole form
\begin{eqnarray}
\bar{F_1}(Q^2)=\frac{\bar{F_1}(0)}{1-Q^{2}/m_{pole}^{2}} \, \, ,
\end{eqnarray}
which works well for light mesons.
Using an $m_{pole}$ of order 
$1.7-1.9$ GeV, which corresponds to the mass of the higher resonances of 
$\phi$, 
we estimate $\bar{F_1}(0)= 0.16 \pm 0.02 $.

With the help of VMD \cite{sakurai} and factorization we can now present the 
amplitude for $B_{s}\rightarrow\gamma\gamma$.
Using the intermediate propagator 
$\frac{-1}{Q^{2}+m_{\phi}^{2}}$ at 
$Q^{2}=0$, the $\phi\rightarrow \gamma$ conversion vertex from the VMD 
mechanism 
\begin{eqnarray}
<0|J_{\mu\,\, em}|\phi(p',\epsilon)>=e Q_{s} f_{\phi}(0) m_{\phi} \epsilon_{\mu} \, \, ,
\end{eqnarray}
and the ${\cal A}(B_{s}\rightarrow\phi\gamma)$ 
amplitude, see eq.~({\ref{amphigam}}),
we get:
\begin{eqnarray}
{\cal A}(B_{s}\rightarrow\phi\gamma\rightarrow\gamma\gamma)
&=&\epsilon_1^{\mu}(k_1) \epsilon_2^{\nu}(k_2)
(A^{+}_{LD_{O_7}} g_{\mu \nu} +
i A^{-}_{LD_{O_7}} \epsilon_{\mu \nu \alpha \beta} k_1^{\alpha}  k_2^{\beta})
\, \, ,
\label{amo7}
\end{eqnarray}
with the CP-even ($A^{+}_{LD_{O_7}}$) and CP-odd ($A^{-}_{LD_{O_7}}$) parts:
\begin{eqnarray}
A^{+}_{LD_{O_7}}&=&2 \chi C m_b \frac{m_{B_s}^2-m_{\phi}^2}{2} \bar{F_1}(0)
\nonumber\\
&=&\sqrt{2} \frac{\alpha_{em} G_F}{\pi} \bar{F_1}(0) f_{\phi}(0) 
\lambda_t \frac{m_b (m_{B_s}^2-m_{\phi}^2)}{3 m_{\phi}} C_{7}^{eff}(\mu)  
 \, \, , \nonumber\\
A^{-}_{LD_{O_7}}&=&2 \chi C m_b \bar{F_1}(0)
\nonumber\\
&=&2 \sqrt{2} \frac{\alpha_{em} G_F}{\pi} \bar{F_1}(0) f_{\phi}(0) 
\lambda_t \frac{m_b}{3 m_{\phi}} C_{7}^{eff}(\mu) \, \, ,
\end{eqnarray}
where $f_{\phi}(0)=0.18$ GeV \cite{terasaki}, $Q_{s}=-1/3$ and
$C$ is defined in eq.~(\ref{const}).
The factor 2 comes from the addition of the diagrams with interchanged photons.
Note, that while for the analysis of the sum rule for 
$B_s \rightarrow \phi \gamma$ we have used 
$f_{\phi}\equiv f_{\phi}(m_{\phi}^2)$, here we take 
into account the suppression in $f_{\phi}(Q^2)$ going from
$Q^2=m_{\phi}^2$ to $Q^2=0$.
We treated the polarization vector $\epsilon^{\phi}$ as transversal and 
replaced $\epsilon \to \epsilon_1, \,\epsilon^{\phi} \to \epsilon_2, 
\, q \to k_1, \, p' \to k_2 $.
The conversion factor $\chi$ is defined as 
\begin{eqnarray}
\chi=-e Q_{s} \frac{f_{\phi}(0)}{m_{\phi}} \, \, .
\label{chi}
\end{eqnarray}

Adding this to the short-distance amplitudes (eq.~(\ref{amplitudes})), 
we obtain the $B_{s}\rightarrow\gamma\gamma$ width 
including the $O_7$-type long-distance effects:
\begin{equation}
\Gamma(B_{s}\rightarrow \gamma \gamma)_{SD+LD_{O_7}}=\frac{1}{32 \pi m_{B_s}} 
(4 |A^{+}+A^{+}_{LD_{O_7}}|^2+\frac{1}{2} m_{B_s}^4|A^{-}+A^{-}_{LD_{O_7}}|^2)
\, \, .
\label{brld}
\end{equation}

\begin{figure}[htb]
\vskip 0.2truein
\centerline{\epsfysize=9.0cm
{\epsffile{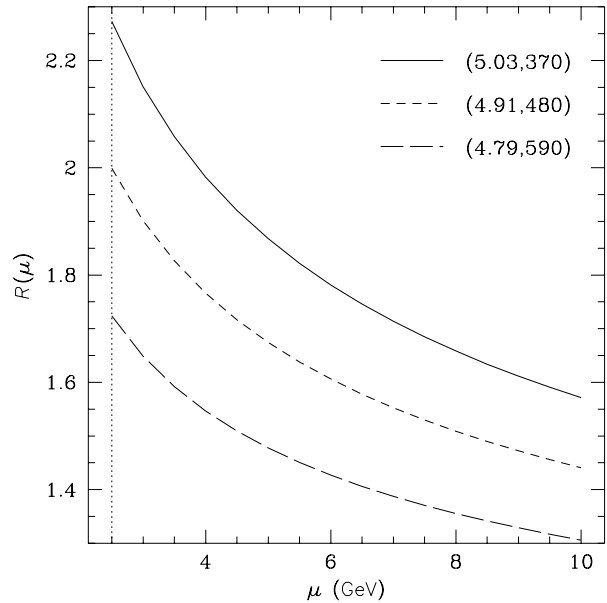}}}
\vskip -0.2truein
\caption[]{Scale dependence of the ratio ${\it R}(\mu)$ defined in 
eq.~(\ref{rat}). The solid, short-dashed and long-dashed lines correspond to
the values $(m_b,\bar{\Lambda}_s)$ in $( {\mbox{GeV}}, {\mbox{MeV}} )$ 
as indicated in the figure.
The dotted line depicts the suggested choice of the scale $\mu$ from
$B \rightarrow X_s \gamma$ studies in NLO \cite{nlo,effham}. 
The parameters used are given in Table \ref{parameters}.}
\label{fig:dependence}
\end{figure}

\section{Numerical estimates}
First we study the leading logarithmic $\mu$-dependence of the ratio
\begin{eqnarray}
{\it R}(\mu)=\frac{\Gamma(B_{s}\rightarrow \gamma \gamma)(\mu)_{SD+LD_{O_7}}}
{\Gamma(B_{s}\rightarrow \gamma \gamma)(m_W)_{SD+LD_{O_7}}} \, \, .
\label{rat}
\end{eqnarray}
In the numerical analysis we neglect the masses of the light quarks.
{}From fig.~\ref{fig:dependence} we find an 
enhancement factor of $1.3-2.3$ relative to the 
lowest  order result obtained by setting $\mu=m_W$, depending on the model 
parameter $(m_b,\bar{\Lambda}_s)$. Varying $\mu$ in the range
$2.5 \, \,{\mbox{GeV}} \leq \mu \leq 10.0 \, \,{\mbox{GeV}}$, gives an 
uncertainty
$\triangle {\it R}/{\it R}(\mu=5\,\,{\mbox{GeV}})\approx \pm (17,19,22) \%$
for $\bar{\Lambda}_s=(590,480,370)$ MeV, respectively.
Here one can argue, that the choice $\mu=\frac{m_b}{2}$ takes into account
effectively the bulk of the NLO correction as suggested by the
NLO calculation for $B \rightarrow X_s \gamma$ \cite{nlo}.

Table~\ref{mums} shows the combined $\mu$ and model parameter dependence of 
the branching ratio 
\begin{equation}
{\cal B}(B_{s} \rightarrow \gamma \gamma)_{SD+LD_{O_7}}=
\frac{\Gamma(B_{s}\rightarrow \gamma \gamma)_{SD+LD_{O_7}}}
 {\Gamma_{tot}(B_s)}\,\,.
\end{equation}
The dependence of the form factor $\bar{F_1}(m_{\phi}^2)$ 
on the $b$ quark mass has been extrapolated from fig.~3 \cite{alibraunsimma}. 
Here $\bar{F_1}(0)= 0.14, 0.15, 0.16$ has been used for 
$m_b=(5.03,4.91,4.79)$ GeV, respectively.
\begin{table}[h]
	\begin{center}
	\begin{tabular}{|c|l|l|l|}
	\hline
	\multicolumn{1}{|c|}{$\mu$}	& 
		\multicolumn{1}{|c|}{$\bar{\Lambda}_s=370$ MeV}	
& 
		\multicolumn{1}{|c|}{$\bar{\Lambda}_s=480$ MeV}
& 
		\multicolumn{1}{|c|}{$\bar{\Lambda}_s=590$ MeV}\\
\multicolumn{1}{|c|}{(GeV)}  
& $m_b=5.03$ GeV & $m_b=4.91$ GeV & $m_b=4.79$ GeV \\
	\hline \hline
$2.5$   & $1.43\cdot 10^{-6}$ & $8.1 \cdot 10^{-7}$ & $5.0 \cdot 10^{-7}$\\
$5.0$   & $1.18\cdot 10^{-6}$ & $6.8 \cdot 10^{-7}$ & $4.3 \cdot 10^{-7}$\\
$10.0$  & $0.99\cdot 10^{-6}$ & $5.9 \cdot 10^{-7}$ & $3.8 \cdot 10^{-7}$\\
\hline 
	\end{tabular}
	\end{center}
\caption{Branching ratio 
${\cal B}(B_{s}\rightarrow \gamma \gamma)_{SD+LD_{O_7}}$ 
for selected values $(m_b,\bar{\Lambda}_s)$ 
and the renormalization scale $\mu$.}
\label{mums}
\end{table}
Qualitatively, the influence of the LD contribution through
$B_{s}\rightarrow\phi\gamma \rightarrow \gamma \gamma$ reduces the width 
because of the destructive interference of the LD $+$ SD contributions.
To quantify this, we define 
\begin{equation}
\kappa \equiv \frac{{\cal B}(B_{s}\rightarrow\gamma\gamma)_{SD+LD_{O_7}}
 -{\cal B}(B_{s}\rightarrow\gamma\gamma)_{SD}}
{{\cal B}(B_{s}\rightarrow\gamma\gamma)_{SD}} \, \, ,
\end{equation}
with $\Gamma(B_{s}\rightarrow \gamma \gamma)_{SD}$ given in 
eq.~(\ref{br}).
We find, that $\kappa$ lies in the range:
\begin{equation}
-15 \% \leq \kappa \leq -27 \% \, \, ,
\end{equation}
depending mainly on $(m_b,\bar{\Lambda}_s)$.

In conclusion, we have reanalysed the decay rate 
$B_{s}\rightarrow\gamma\gamma$ in the SM. 
We included the leading logarithmic QCD corrections and investigated the 
influence of the LD-contributions due to the chain
$B_{s}\rightarrow\phi\gamma \rightarrow\gamma\gamma $. 
Depending on $\bar{\Lambda}_s$, the LD-contributions become sizeable.
Other possible LD contributions may also arise from the $O_2$-type transitions.

The decay rate of $B_s \rightarrow \gamma \gamma$ depends sensitively on 
$(m_b,\bar{\Lambda}_s)$ and $\mu$. 
Fixing $\mu$ to $\mu=\frac{m_b}{2}$ as suggested by the NLO
calculation of $B \rightarrow X_s \gamma$ and varying 
the model parameter $(m_b,\bar{\Lambda}_s)$ (see Table~\ref{mums}), 
we find that the branching ratio 
${\cal B}(B_s \rightarrow \gamma \gamma)_{SD+LD_{O_7}}$ 
is uncertain by a large factor
\begin{equation}
0.5 \cdot 10^{-6} \leq
{\cal B}(B_s \rightarrow \gamma \gamma)_{SD+LD_{O_7}} 
\leq 1.4 \cdot 10^{-6} \, \, .
\end{equation}
Improving this requires NLO calculation in the decay rate 
$B_s \rightarrow \gamma \gamma$.
With the choice of 
$(m_b,\bar{\Lambda}_s)=(5.03 \,\mbox{GeV},370 \, \mbox{MeV})$, 
the resulting branching ratio 
$(1.4 \cdot 10^{-6})$ is substantially larger than what has been stated in 
the literature. 
The present best limit on the decay $B_s \rightarrow \gamma \gamma$
is \cite{L3} 
\begin{equation}
{\cal B}(B_s \rightarrow \gamma \gamma) < 1.48 \cdot 10^{-4} \, \, ,
\nonumber
\end{equation}
which is still a factor $\approx 100-300$ away from the estimates given here.

\noindent
Note added: Recently, the leading logarithmic QCD corrections for the
short-distance part of the decay $B_s \rightarrow \gamma \gamma$ 
have also been calculated by Chang et al. \cite{yaonew}.
They derived the decay rate with the full set of operators $O_{1 \dots 8}$ 
and we agree with their analytical expression.
%
%
Our model to incorporate the bound state effects in the $B_s$ meson is 
inspired by HQET, resulting in the parameters $(m_b,\bar{\Lambda}_s)$.
The strong parametric dependence
of the decay rate $\Gamma(B_s \rightarrow \gamma \gamma)$ on 
$(m_b,\bar{\Lambda}_s)$ and on $\mu$ has been studied by us; 
Chang et al. \cite{yaonew} fix $\Lambda=m_{B_s}-m_b$, using the naive 
constituent quark model, and set $\mu=m_b$. 
We emphasize here that the decay rate is sensitive to both of these 
parameters and requires further theoretical investigation.

\bigskip
\noindent
{\Large \bf Acknowledgements}

We thank Ahmed Ali for helpful discussions and input and Christoph Greub for 
discussions, especially on QCD corrections. 
E. O. I. would also like to thank the DESY theory group for the 
warm hospitality during his stay in Hamburg. The work of E. O. I. was 
supported by the Turkish Scientific and Technical Research Council (TUBITAK).

\bigskip
\bigskip
%

%

\begin{thebibliography}{1}
\bibitem{cleo}
    M. S. Alam et al., (CLEO Collaboration), Phys. Rev. Lett. {\bf 74} (1995)
2885.
\bibitem{nlo}
        C. Greub, T. Hurth and D. Wyler, Phys. Lett.B  {\bf 380} (1996) 385;
 Phys. Rev. D {\bf 54} (1996) 3350; 
 K. Chetyrkin, M. Misiak and M. M\"unz, hep-ph/9612313.
\bibitem{effham}
	A. Ali and C. Greub, Z. Phys. C {\bf 49} (1991) 431; \\
  A. J. Buras et al., Nucl. Phys. B {\bf 424} (1994) 374. 
\bibitem{yao}
   G.-L. Lin, J. Liu and Y.-P. Yao, Phys. Rev. Lett. {\bf 64} (1990) 1498; \\
   G.-L. Lin, J. Liu and Y.-P. Yao, Phys. Rev. D {\bf 42} (1990) 2314.
\bibitem{simma}
        H. Simma and D. Wyler, Nucl. Phys. B {\bf 344} (1990) 283.
\bibitem{aliev}
        T. M. Aliev and G. Turan, Phys. Rev. D {\bf 48} (1993) 1176.
\bibitem{grinstein}
  B. Grinstein, R. Springer and M. Wise, Phys. Lett. B {\bf 202} (1988) 138. 
\bibitem{MW}
  A. Manohar and M. B. Wise,
        Phys. Rev. D {\bf 49} (1994) 1310.
\bibitem{terasaki} K. Terasaki, Nuovo Cim. Vol. {\bf 66} A, No. {\bf 4} (1981) 475.
\bibitem{alibraunsimma}
A. Ali, V. M. Braun and H. Simma, Z. Phys. C {\bf 63} (1994) 437. 
\bibitem{gremm} 
 M. Gremm, A. Kapustin, Z. Ligeti and M.B. Wise, Phys. Rev. Lett.
        {\bf 77} (1996) 20.
\bibitem{PDG} R. M. Barnett et al., Review of Particle Properties, Phys Rev D
{\bf 54} (1996) 1.
\bibitem{alibraun} A. Ali and V. M. Braun, Phys. Lett. B {\bf 359} (1995) 223.
\bibitem{colangelo} P. Colangelo et al., Phys. Lett. B {\bf 317} (1993) 183.
\bibitem{mzero} V.M.Belyaev, B.L.Ioffe, Sov. JETP {\bf 83} (1982) 876.
\bibitem{sbars} B.L.Ioffe, A.V.Smilga, Phys.Let. B {\bf 133} (1985) 436.
\bibitem{shifman} M. A. Shifman, A. I. Vainshtein and V. I. Zakharov, 
 Nucl. Phys. B {\bf 147} (1979) 385.
\bibitem{chernyak}
  V. L. Chernyak and A. R. Zhitnitsky, Phys. Rep. {\bf 112} ( 1984) 173.
\bibitem{sakurai} J. J. Sakurai, Currents and Mesons (University of Chicago
Press, Chicago, 1969); \\ 
 E. Golowich and S. Pakvasa, Phys. Rev. D {\bf 51} (1995) 1215; \\
N. G. Deshpande, Xiao-Gang He and J. Trampetic, Phys. Lett. B {\bf 367} (1996) 
362. 
\bibitem{L3}
 M. Acciarri et al. (L3 Collaboration), Phys. Lett. B {\bf 363} (1995) 137.
%
\bibitem{yaonew}
Chia-Hung V. Chang, Guey-Lin Lin and York-Peng Yao, hep-ph/9705345.
\end{thebibliography}
\end{document}